\documentclass[prb,reprint,floatfix,]{revtex4-1}
\usepackage{siunitx}
\usepackage{graphicx}
\usepackage{xcolor}
\usepackage{amsmath}

\newcommand{\mrx}{Mn$_2$Ru$_x$Ga}
\newcommand{\mrg}{MRG}
\newcommand{\tcmp}{T_M}
\newcommand{\geff}{g_{\text{eff}}}

\begin{document}
\author{G. Bonfiglio}
\affiliation{Radboud University, Institute for Molecules and Materials, 6525 AJ Nijmegen, The Netherlands}
\author{K. Rode}
\author{K. Siewerska}
\author{J. Besbas}
\author{G. Y. P. Atcheson}
\author{P. Stamenov}
\author{J.M.D.~Coey}
\affiliation{CRANN, AMBER and School of Physics, Trinity College Dublin, Ireland}
\author{A.V. Kimel}
\author{Th. Rasing}
\author{A. Kirilyuk}
\affiliation{Radboud University, Institute for Molecules and Materials, 6525 AJ Nijmegen, The Netherlands}
\affiliation{FELIX Laboratory, Radboud University, Toernooiveld 7c, 6525 ED Nijmegen, The Netherlands}

\title{Magnetisation dynamics of the compensated ferrimagnet \mrx}

\begin{abstract}
  Here we study both static and time-resolved dynamic magnetic properties of the compensated
  ferrimagnet \mrx\ from room temperature down to \SI{10}{\kelvin}, thus crossing
  the magnetic compensation temperature $\tcmp$. The behaviour
  is analysed with a model of a simple collinear ferrimagnet with uniaxial anisotropy and
  site-specific gyromagnetic ratios. We find a maximum zero-applied-field
  resonance frequency of $\sim \SI{160}{\giga\hertz}$ and a low intrinsic
  Gilbert damping $\alpha \sim 0.02$, making it a very attractive candidate for various spintronic applications.
\end{abstract}

\maketitle

\section{Introduction}
\label{sec:intro}
Antiferromagnets (AFM) and compensated ferrimagnets (FiM) have attracted a lot of attention
over the last decade due to their potential use in spin electronics\cite{Shick2010,Caretta2018}.
Due to their lack of a net magnetic moment, they are insensitive to
external fields and create no demagnetising fields of their own. In addition, their spin
dynamics reach much higher frequencies than those of their ferromagnetic (FM)
counterparts due to the contribution of the exchange energy in the magnetic
free energy\cite{Gomonay2014}.

Despite these clear advantages, AFMs are scarcely used apart from uni-directional exchange biasing relatively
in spin electronic applications. This is because the lack of net moment also implies that there is no direct way to manipulate their magnetic state. Furthermore, detecting their magnetic state
is also complicated and is usually possible only by neutron diffraction
measurements\cite{Shull1949}, or through interaction with an adjacent FM layer\cite{Jungwirth2016}.

Compensated, metallic FiMs provide an interesting alternative as they
combine the high-speed advantages of AFMs with those of FMs, namely, the ease to manipulate their magnetic state.
Furthermore, it has been shown that such materials are good candidates for the emerging field of All-Optical Switching (AOS) in which the magnetic state is solely controlled by a fast laser pulse \cite{Stanciu2007,Mangin2014,Banerjee2019}.
A compensated, half-metallic ferrimagnet was first envisaged by \citet{VanLeuken1995}. In their model two magnetic ions in
crystallographically different positions couple antiferromagnetically and
perfectly compensate each-other, but only one of the two contributes to the
states at the Fermi energy responsible for electronic transport. The first
experimental realisation of this, \mrx~(\mrg), was provided by \citet{Kurt2014}.

\mrg\ crystallises in the $XA$ Heusler structure, space group $F\bar{4}3m$, with Mn on the $4a$ and $4c$
sites\cite{Betto2015}. Substrate-induced bi-axial strain imposes a slight
tetragonal distortion, which leads to perpendicular magnetic anisotropy. Due to
the different local environment of the two sublattices, the temperature
dependence of their magnetic moments differ, and perfect compensation is
therefore obtained at a specific temperature $\tcmp$ that depends on the Ru
concentration $x$ and the degree of biaxial strain. It was previously shown
that \mrg\ exhibits properties usually associated with FMs: a large anomalous
Hall angle\cite{Thiyagarajah2015}, that depends only on the magnetisation of
the $4c$ magnetic sublattice\cite{Fowley2018}; tunnel magnetoresistance
(TMR) of \SI{40}{\percent}, a signature of its high spin
polarisation\cite{Zic2016}, was observed in magnetic tunnel junctions (MTJs)
based on \mrg\cite{Borisov2016}; and a clear magneto-optical Kerr effect and domain structure, even in the
absence of a net moment\cite{Fleischer2018,8478820}. Strong exchange bias of a CoFeB
layer by exchange coupling with \mrg\ through a Hf spacer
layer\cite{Borisov2017}, as well as single-layer spin-orbit
torque\cite{troncoso2018antiferromagnetic,lenne2019giant} showed that \mrg\
combined the qualities of FMs and AFMs in spin electronic devices.

The spin dynamics in materials where two distinct sublattices are subject to
differing internal fields (exchange, anisotropy, \ldots) is much richer than
that of a simple FM, as previously demonstrated by the obersvation of single-pulse all-optical switching in amorphous GdFeCo\cite{Stanciu2006,Radu2011} and very recently in \mrg\cite{Banerjee2019}. Given that the magnetisation of \mrg\ is small, escpecially close to the compensation point, and the related frequency is high, normal ferromagnetic resonance (FMR) spectroscopy is unsuited to study their properties. Therefore, we used the all-optical pump-probe technique to characterize the resonance frequencies at different temperatures in vicinity of the magnetic compensation point. This, together with the simulation of FMR, make it possible to determine the effective g-factors, the anisotropy constants and their evolution across the compensation point.
We found, in particular, that our ferrimagnetic half-metallic Heusler alloy has resonance frequency up to 160 GHz at zero-field and a relatively low Gilbert damping.

\section{Experimental details}
\label{sec:expdetail}

\begin{figure}
  \centering
  \includegraphics[width=\columnwidth]{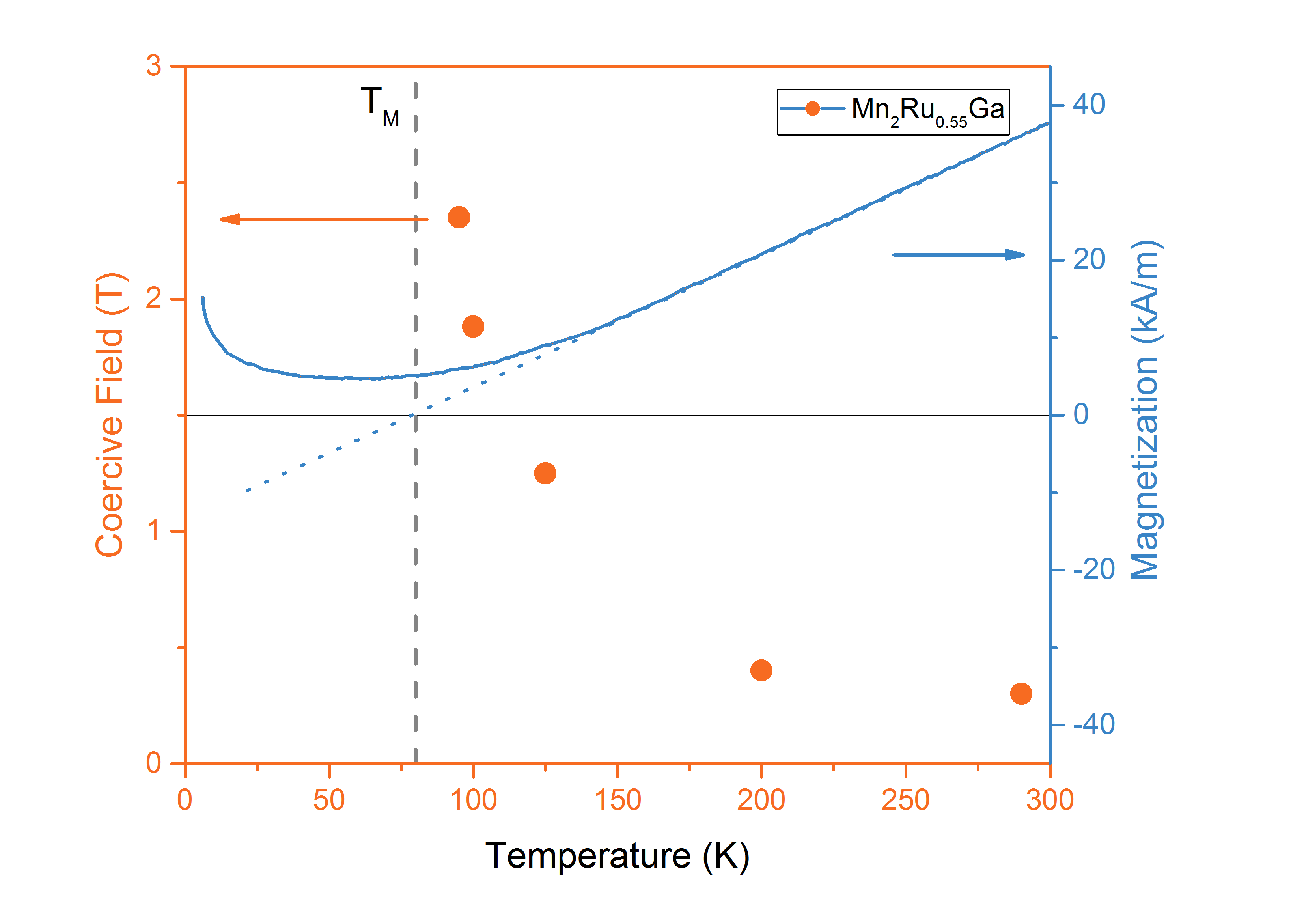}
  \caption{Net moment measured by magnetometry and coercive field measured by
    static Faraday effect. The upturn of the net moment below $T \sim \SI{50}{\kelvin}$
    is due to paramagnetic impurities in the MgO substrate. $\tcmp$ is indicated by
    the vertical dotted line. As expected the maximum available applied field
    $\mu_0 H = \SI{7}{\tesla}$ is insufficient to switch the magnetisation close to
    $\tcmp$. }
  \label{fig:mt-coercivity}
\end{figure}
Thin film samples of \mrg\ were grown in a `Shamrock' sputter deposition
cluster with a base pressure of \SI{2e-8}{Torr} on MgO (001) substrates.
Further information on sample deposition can be found elsewhere\cite{Betto2016}. The substrates were kept at \SI{250}{\degreeCelsius}, and a protective $\sim \SI{3}{\nano\meter}$ layer of
aluminium oxide was added at room temperature. Here we focus on
a \SI{53}{\nano\metre} thick sample with $x = 0.55$, leading to $\tcmp \approx
\SI{80}{\kelvin}$ as determined by SQUID magnetometry using a Quantum Design
\SI{5}{\tesla} MPMS system (see \figurename~\ref{fig:mt-coercivity}). We are
able to study the magneto-optical properties both above and below $\tcmp$.

The magnetisation dynamics was investigated using an all-optical two-colour pump-probe scheme in
a Faraday geometry inside a $\mu_0 H_{\text{max}} = \SI{7}{\tesla}$
superconducting coil-cryostat assembly.  Both pump and probe were produced by a
Ti:sapphire femtosecond pulsed laser with a central wavelength of
\SI{800}{\nano\metre}, a pulse width of \SI{40}{\femto\second} and a repetition
rate of \SI{1}{\kilo\hertz}. After splitting the beam in two, the
high-intensity one was doubled in frequency by a BBO crystal (giving
$\lambda=\SI{400}{\nano\metre}$) and then used as the pump while the lower
intensity \SI{800}{\nano\metre} beam acted as the probe pulse. The time delay
between the two was adjusted by a mechanical delay stage. The pump was
then modulated by a synchronised mechanical chopper at \SI{500}{\hertz} to
improve the signal to noise ratio by lock-in detection. Both pump and probe
beams were linearly polarized, and with spot sizes on the sample of
\SIlist{150;70}{\micro\metre}, respectively. The pump pulse hit the sample at
an incidence angle of $\approx \SI{10}{\degree}$.  After interaction with the
sample, we split the probe beam in two orthogonally polarized parts using a
Wollaston prism and detect the changes in transmission and rotation by
calculating the sum and the difference in intensity of the two signals.

The external field was applied at \SI{75}{\degree} to the easy axis of
magnetization thus tilting the magnetisation away from the axis. Upon
interaction with the pump beam the magnetisation is momentarily drastically
changed\cite{Koopmans2005} and we monitor its return to the initial
configuration via remagnetisation and then precession through the time dependent Faraday
effect on the probe pulse.

The static magneto-optical properties were examined in the same
cryostat/magnet assembly.

\section{Results \& discussion}
\label{sec:results}

\subsection{Static magnetic properties}
\label{ssec:faraday_squid_hall}
\begin{figure}
  \centering
  \includegraphics[width=\columnwidth]{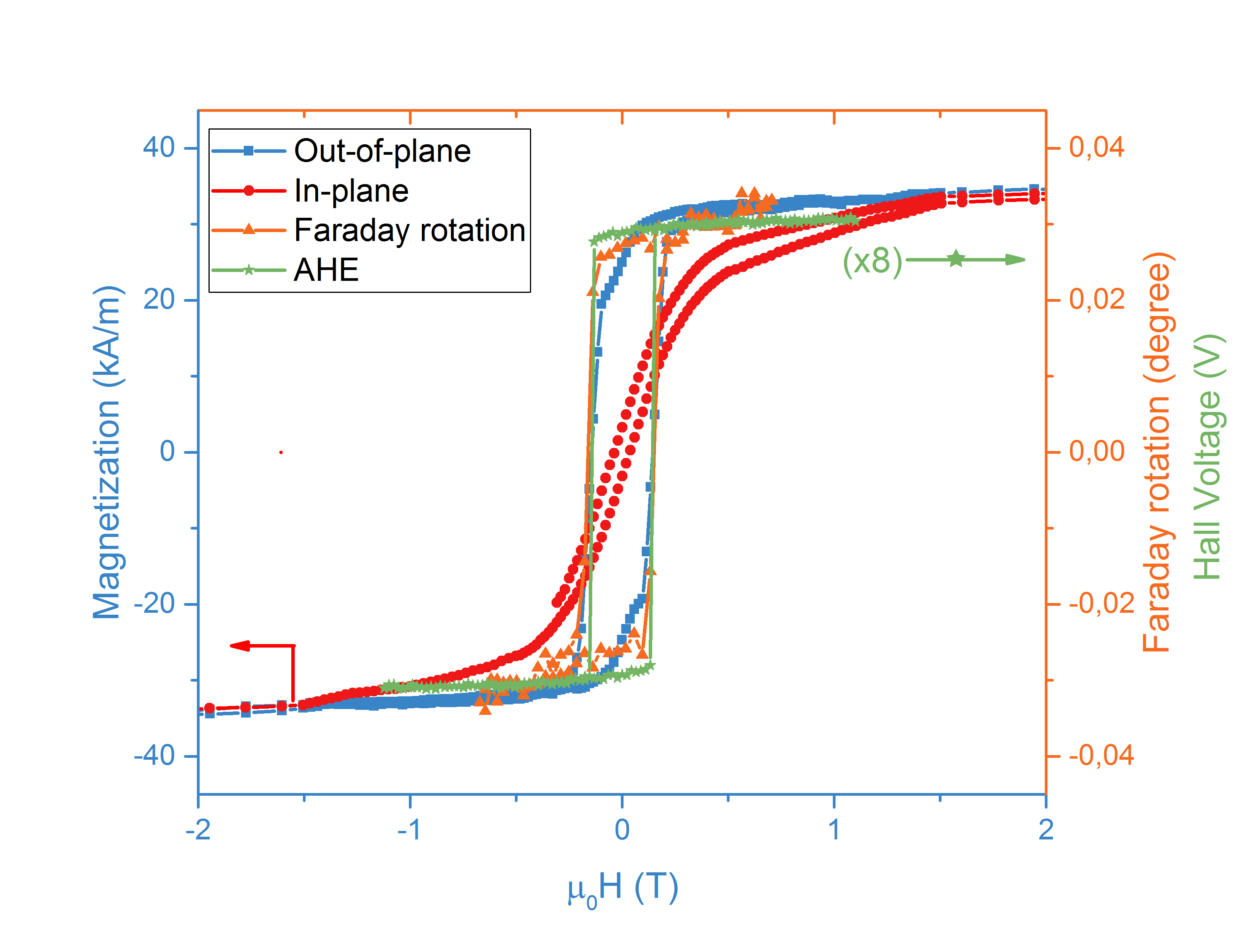}
  \caption{Comparison of hysteresis
    loops obtained by Faraday, AHE, and magnetometry recorded at room
    temperature. The two former were recorded with the applied field
    perpendicular to the sample surface, while for the latter we show results
    for both field applied parallel and perpendicular to the sample.}
  \label{fig:static_comparison}
\end{figure}
We first focus on the static magnetic properties as observed by the Faraday
effect, and compare them to what is inferred from magnetometry and the anomalous
Hall effect. In \figurename~\ref{fig:static_comparison} we present magnetic
hysteresis loops as recorded using the three techniques. Due to the half metallic
nature of the sample, the magnetotransport properties depend only on the
$4c$ sublattice. As the main contribution to the \mrg\ dielectric tensor in the visible and near infrared arises from the Drude tail\cite{Fleischer2018}, both AHE and Faraday effect
probe essentially the same properties (mainly the spin polarised conduction
band of \mrg), hence we observe overlapping loops for the two
techniques. Magnetometry, on the other hand, measures the net moment, or to be
precise the \emph{small difference} between two large sublattice moments. The $4a$
sublattice, which is insignificant for AHE and Faraday here contributes on
equal footing. \figurename~\ref{fig:static_comparison} shows a clear
difference in shape between the magnetometry loop and the AHE or
Faraday loops. We highlight here that the apparent `soft' contribution that shows switching
close to zero applied field, is not a secondary magnetic phase, but a signature
of the small differences in the field-behaviour of the two sublattices. We also
note that this behaviour is a result of the non-collinear magnetic order of
\mrg . A complete analysis of the dynamic properties therefore requires
knowledge of the anisotropy constants on \emph{both} sublattices as well as the
(at least) three intra and inter sublattice exchange constants. Such an
analysis is beyond the scope of this article, and we limit our analysis to the
simplest model of a single, effective uniaxial anisotropy constant
$K_u$ in the exchange approximation of the ferrimagnet.

\subsection{Dynamic properties}
\label{ssec:fmr}
We now turn to the time-resolved Faraday effect and spin dynamics.
Time-resolved Faraday effect data were recorded at five different temperatures
\SIlist{10;50;100;200;290}{\kelvin}, with applied fields ranging from 
\SIrange{1}{7}{\tesla}. 

\begin{figure}
  \centering
  \includegraphics[width=\columnwidth]{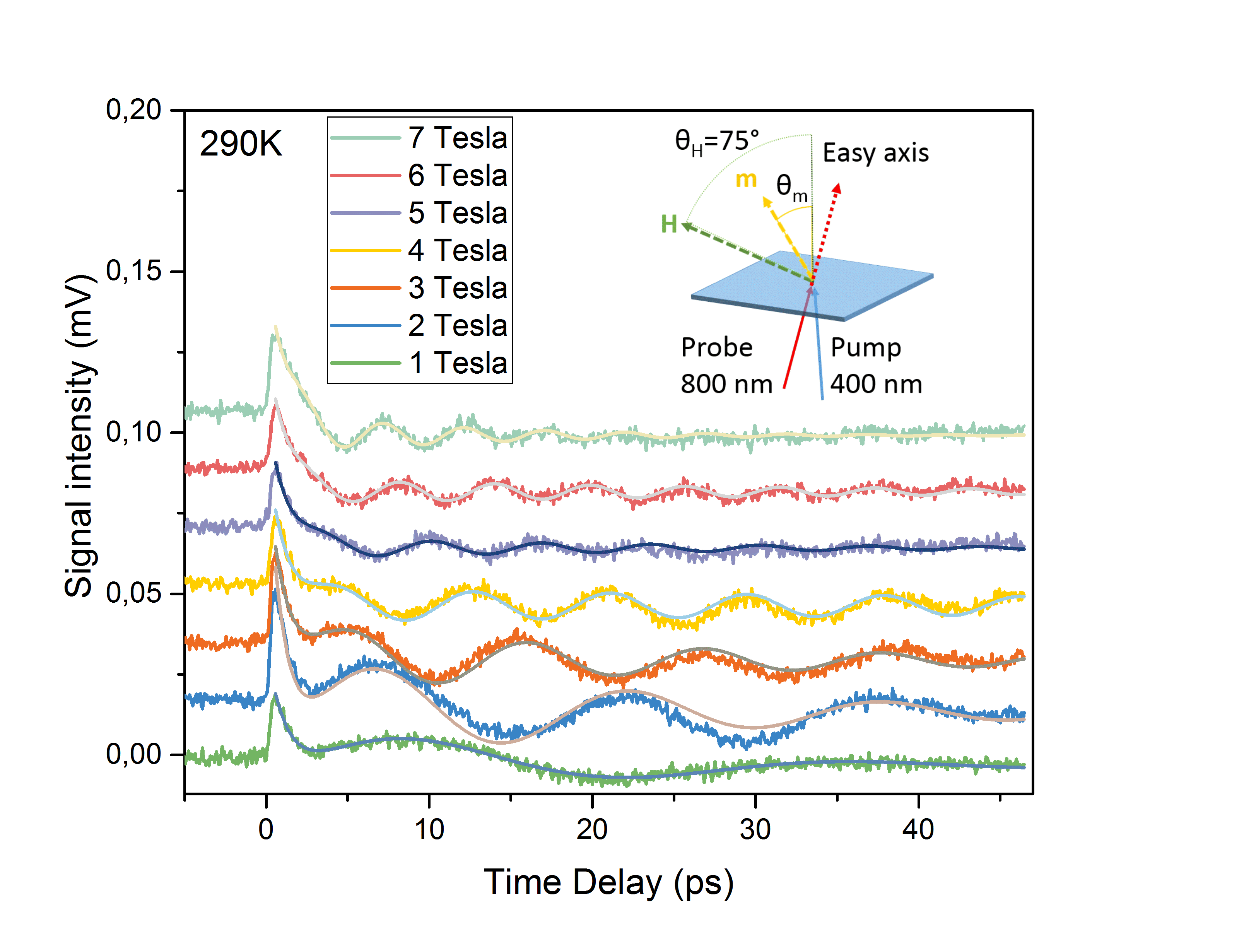}
  \caption{Time resolved Faraday effect recorded at $T = \SI{290}{\kelvin}$ in
    applied fields ranging from \SIrange{1}{7}{\tesla}. After the initial
    demagnetisation seen as a sharp increase in the signal at $t \sim
    \SI{0}{\pico\second}$, magnetisation is recovered and followed by
    precession around the effective field until fully damped. The lines are fits
    to the data. The inset shows the experimental geometry further detailed in
    the main text.}
  \label{fig:trfaraday290}
\end{figure}
\figurename~\ref{fig:trfaraday290} shows the field-dependence of the Faraday effect
as a function of the delay between the pump and the probe pulses, recorded
at $T = \SI{290}{\kelvin}$. Negative delay indicates the probe is hitting the
sample before the pump. After the initial demagnetisation, the magnetisation
recovers and starts precessing around the effective field which is determined
by the anisotropy and the applied field. The solid lines in
\figurename~\ref{fig:trfaraday290} are fits to the data to extract the period
and the damping of the precession in each case. The fitting model was an
exponentially damped sinusoid with a phase offset. We note that the apparent
evolution of the amplitude and phase with changing applied magnetic field is
due to the quasi-resonance of the spectrum of the precessional motion with the
low-frequency components of the convolution between the envelope of the probe
pulse and the physical relaxation of the system. The latter include both
electron-electron and electron-lattice effects. A rudimentary model based on a
classical oscillator successfully reproduces the main features of the amplitude
and phase observed.

In two-sublattice FiMs, the gyromagnetic ratios of the two sublattices are not
necessarily the same. This is particularly obvious in rare-earth/transition
metal alloys, and is also the case for \mrg\, despite the two sublattices being
chemically similar; they are both Mn. Due to the different local environment
however, the degree of charge transfer for the two differs. This leads to two
characteristic temperatures, a first $\tcmp$ where the magnetic moments
compensate, and a second $T_A$ where the angular momenta compensate. It can be
shown that for the ferromagnetic mode, the effective gyromagnetic ratio
$\gamma_{\text{eff}}$ can then be written\cite{Wangsness153}
\begin{equation}
  \gamma_{\text{eff}} = \frac{M_{4c}(T) - M_{4a}(T)}{M_{4c}(T)/\gamma_{4c} - M_{4a}(T)/\gamma_{4a}}
  \label{eq:gammaeff}
\end{equation}
subscript $i = 4a, 4c$ denotes sublattice $i$,
$M_i(T)$ the temperature-dependent magnetisation, and $\gamma_i$ the
sublattice-specific gyromagnetic ratio. $\gamma_{\text{eff}}$ is related to the
effective $g$-factor
\begin{equation}
  \geff = \gamma_{\text{eff}}\frac{h}{\mu_B}
  \label{eq:geff}
\end{equation}
where $h$ is the Planck constant and $\mu_B$ the Bohr magneton.

The frequency of the precession is determined by the effective field, which can
be inferred from the derivative of the magnetic free energy density with respect
to $\mathbf{M}$. For an external field applied at a given fixed angle with
respect to the easy axis this leads to the Smit-Beljers formula\cite{Smit1955}
\begin{equation}
  \omega_{FMR}=\gamma_{\text{eff}} \sqrt{\frac{1}{M_s^2 sin^2\phi}
  \left[\frac{\delta^2 E}{\delta\theta^2}\frac{\delta^2
E}{\delta\phi^2}-\left(\frac{\delta^2 E}{\delta\theta\delta\phi}\right)^2
\right]}
  \label{eq:smitbeljers}
\end{equation}
where $\theta$ and $\phi$ are the polar and azimuthal angles of the
magnetisation vector, and $E$ the magnetic free energy density
\begin{equation}
  E = - \mu_0 \mathbf{H} \cdot \mathbf{M} + K_u \sin^2{\theta} + \mu_0 M_s^2
  \cos^2{\theta} / 2
  \label{eq:freenrj}
\end{equation}
where the terms correspond to the Zeeman, anisotropy and demagnetising
energies, respectively, and $M_s$ is the net saturation magnetisation. It should be mentioned that the magnetic anisotropy constant $ K_{u}$ is related to $M$, which is being considered constant in magnitude, via $K_u=\beta \mu_0 M_{s}^{2}/2$, $\beta$ a dimensionless parameter.   

\begin{figure}
  \centering
  \includegraphics[width=\columnwidth]{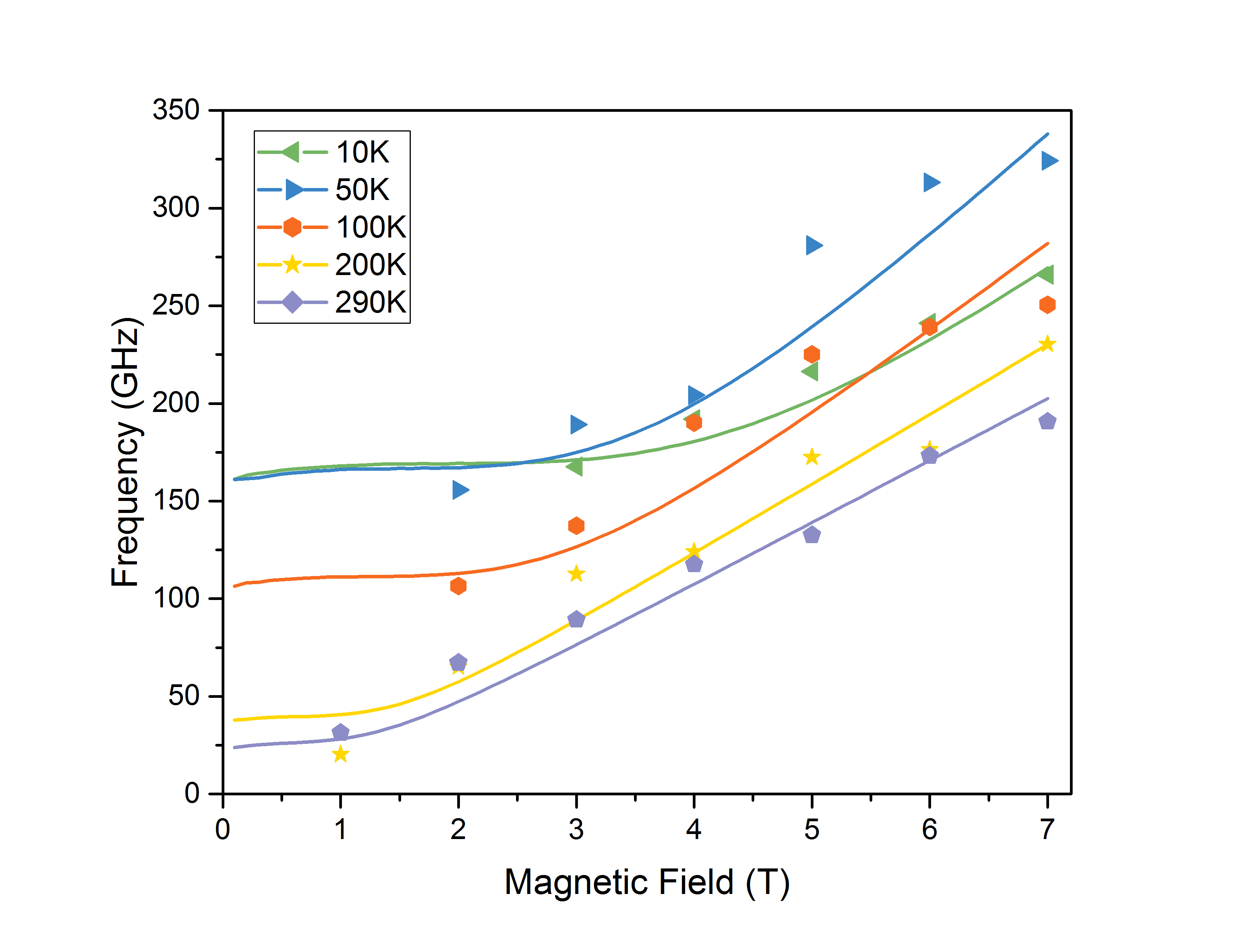}
  \caption{Observed precession frequency as a function of the applied field for
  various temperatures. The solid lines are fits to the data as described in
  the main text.}
  \label{fig:fiveTffit}
\end{figure}

Based on Eqs.~\eqref{eq:gammaeff}~through~\eqref{eq:freenrj} we
fit our entire data set with $\gamma_{\text{eff}}$ and $K_u$ as the only free
parameters. The experimental data and the associated fits are shown as points
and solid lines in \figurename~\ref{fig:fiveTffit}. At all temperatures our
simple model with one effective gyromagnetic ratio $\gamma_{\text{eff}}$ and a
single uniaxial anisotropy parameter $K_u$ reproduces the experimental data
reasonably well. The model systematically underestimates the
resonance frequency for intermediate fields, with the point of maximum disagreement
increasing with decreasing temperature. We speculate this is due to
the use of a simple uniaxial anisotropy in the free energy (see Eq.~\ref{eq:freenrj}), while the real situation is more likely to be better represented as a sperimagnet. In particular, the non-collinear nature of \mrg\ that leads to a deviation from
\SI{180}{\degree} of the angle between the two sublattice magnetisations, depending on the applied field and temperature.

\begin{figure}
  \centering
  \includegraphics[width=\columnwidth]{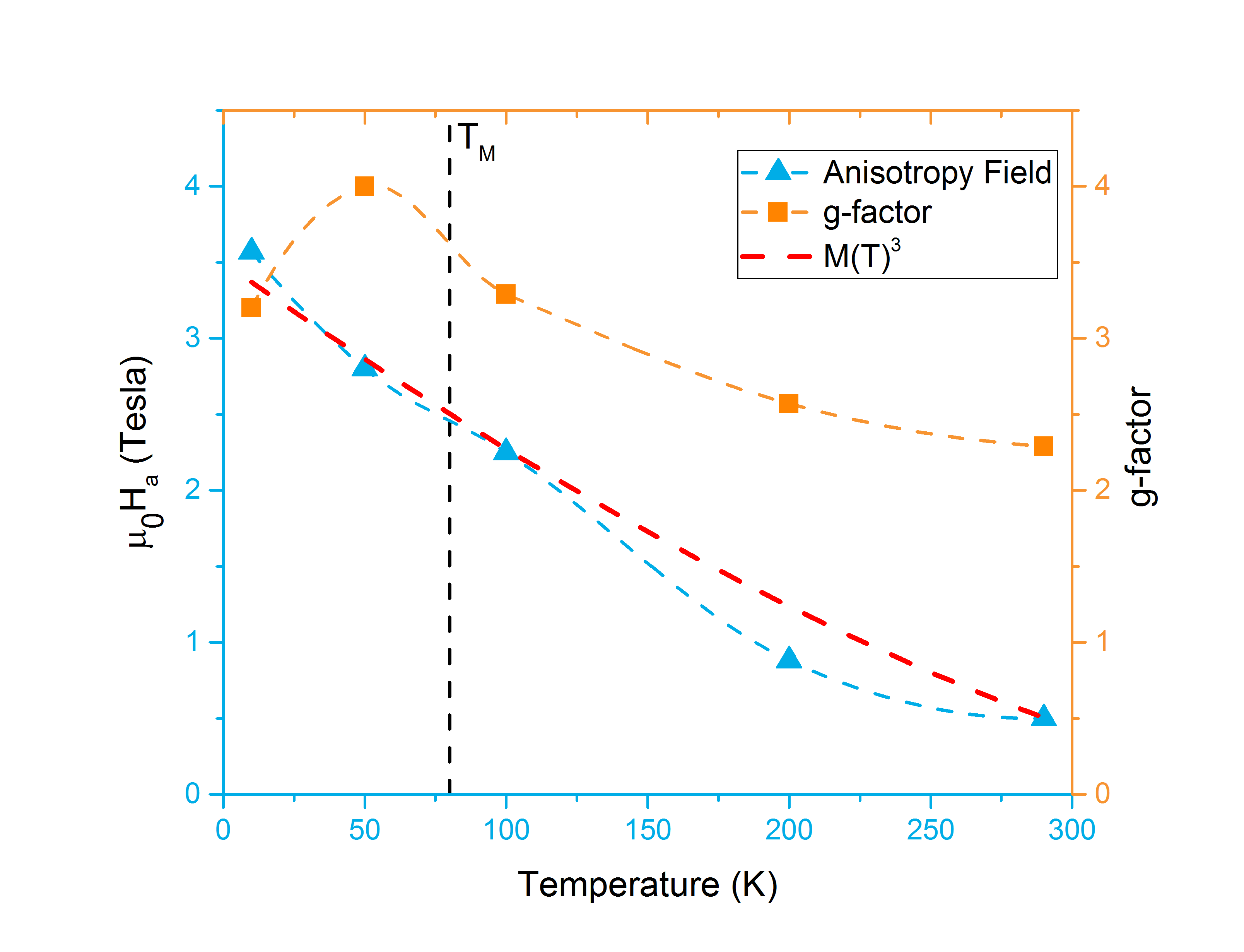}
  \caption{Effective $g$-factor, $\geff$, and the anisotropy field as
  determined by time-resolved Faraday effect. $\geff$, orange squares, increases from near the free electron value of 2 to 4 just below $\tcmp$, while the anisotropy field, blue triangles, increases near-linearly with decreasing temperature. A $M^3$ fit, red dashes line, of the anisotropy behaviour shows the almost-metallic origin of it, indicating the dominant character of the 4c sublattice.}
  \label{fig:geffandhk}
\end{figure}
From the fits in \figurename~\ref{fig:fiveTffit} we infer the values of $\geff$
and the anisotropy field $\mu_0 H_a = {}^{2K_u}\!/\!_{M_s}$. The result is
shown in \figurename~\ref{fig:geffandhk}. The anisotropy field is monotonically
increasing with decreasing temperature as the magnetisation of the $4c$
sublattice increases in the same temperature range. We highlight here the
advantage of determining this field through time-resolved magneto-optics as
opposed to static magnetometry and optics. Indeed the anisotropy field as seen by
static methods is sensitive to the combination of anisotropy and the \emph{net}
magnetic moment, as illustrated in \figurename~\ref{fig:mt-coercivity}, where the coercive
field diverges as $T \rightarrow \tcmp$. In statics one would expect a
divergence of the anisotropy field at the same temperature. The time-resolved
methods however distinguish between the net and the sublattice moments, hence
better reflecting the evolution of the intrinsic material properties of the
ferrimagnet.

The temperature dependence of the anisotropy constants was a matter for discussion for many years\cite{Callen1966,vonsovskii1974magnetism}. Written in spherical harmonics the $3d$ anisotropy can be expressed as, $ k_{2}Y_{2}^0(\theta)+k_{4}Y_{4}^0(\theta)$ ~\cite{Farle1998} where $k_{2}\propto M(T)^3$ and $k_{4}\propto M(T)^{10}$. The experimental measured anisotropy is then,  $K_{2}(T)=ak_{2}(T)+bk_{4}(T)$, with $a$ and $b$ the contributions of the respective spherical harmonics.

\figurename~\ref{fig:geffandhk} shows that a reasonable fit of our data is obtained with $M(T)^3$ which means, first, that the contribution of the $4^{th}$ order harmonic can be neglected, and second, that the contribution of the $\tcmp$ and $2^{nd}$ sublattice is negligible, indicating the dominant character of the 4c sublattice.

In addition, we should note here that the high frequency exchange mode was never observed on our experiments. While far from $\tcmp$ its frequency might be too high to be observable, in the vicinity of $\tcmp$, in contrast, its frequency is expected to be in the detection range. Moreover, given the different electronic structure of the two sublattices, it is expected that the laser pulse should selectively excite the sublattice 4c, and therefore lead to the effective excitation of the exchange mode. We argue that it is the non-collinearity of the sublattices (see section III A) that smears out the coherent precession at high frequencies.

The effective gyromagnetic ratio, $\geff$, shows a non-monotonic
behaviour. It increases with decreasing $T$ towards $\tcmp$, reaching a maximum
at about \SI{50}{\kelvin} before decreasing again at $T =\SI{10}{\kelvin}$. We alluded
above to the difference between the magnetic and the angular momenta
compensation temperatures. We expect that $\geff$ reaches a maximum when
$T = T_A$\cite{gurevich1996}, here between the measurement at $T =\SI{50}{\kelvin}$ and the
magnetic compensation temperature $\tcmp\approx\SI{80}{\kelvin}$.

From XMCD data\cite{Betto2015}, we could estimate spin and orbital moment components of the magnetic moments of the two sublattices, what allowed us to derive the effective g-factors for the sublattices as $g_{4a}=2.05$ and $g_{4c}=2.00$. In this case we expect the angular momentum compensation temperature $T_{A}$ to be below $\tcmp$, opposite to what is observed for GdFeCo\cite{Stanciu2006}. Given this small difference however, $T_{A}$ and $\tcmp$ are expected to be rather close to each other, consistent with the limited increase of $\geff$ across the compensation points.

We turn finally to the damping of the precessional motion of $\mathbf{M}$ around
the effective field $\mu_0 \mathbf{H_{\text{eff}}}$. Damping is usually described via
the dimensionless parameter $\alpha$ in the Landau-Lifshiz-Gilbert equation,
and it is a measure of the dissipation of magnetic energy in the system. In this
model, $\alpha$ is a scalar constant and the observed broadening in the time
domain is therefore a linear function of the frequency of precession\cite{Malinowski2009,Liu2010,Schellekens2013}. We infer
$\alpha '$, the total damping, from our fits of the time-resolved Faraday
effect as $\alpha ' = \left( \tau_d \right)^{-1} $, where $\tau_d$ is
the decay time of the fits. We then, for each temperature, plot $\alpha '$ as a
function of the observed frequency and regress the data using a
straight line fit. The intrinsic $\alpha$ is the slope of this line, while
the intercept represents the anisotropic broadening.

\begin{figure}
  \centering
  \includegraphics[width=\columnwidth]{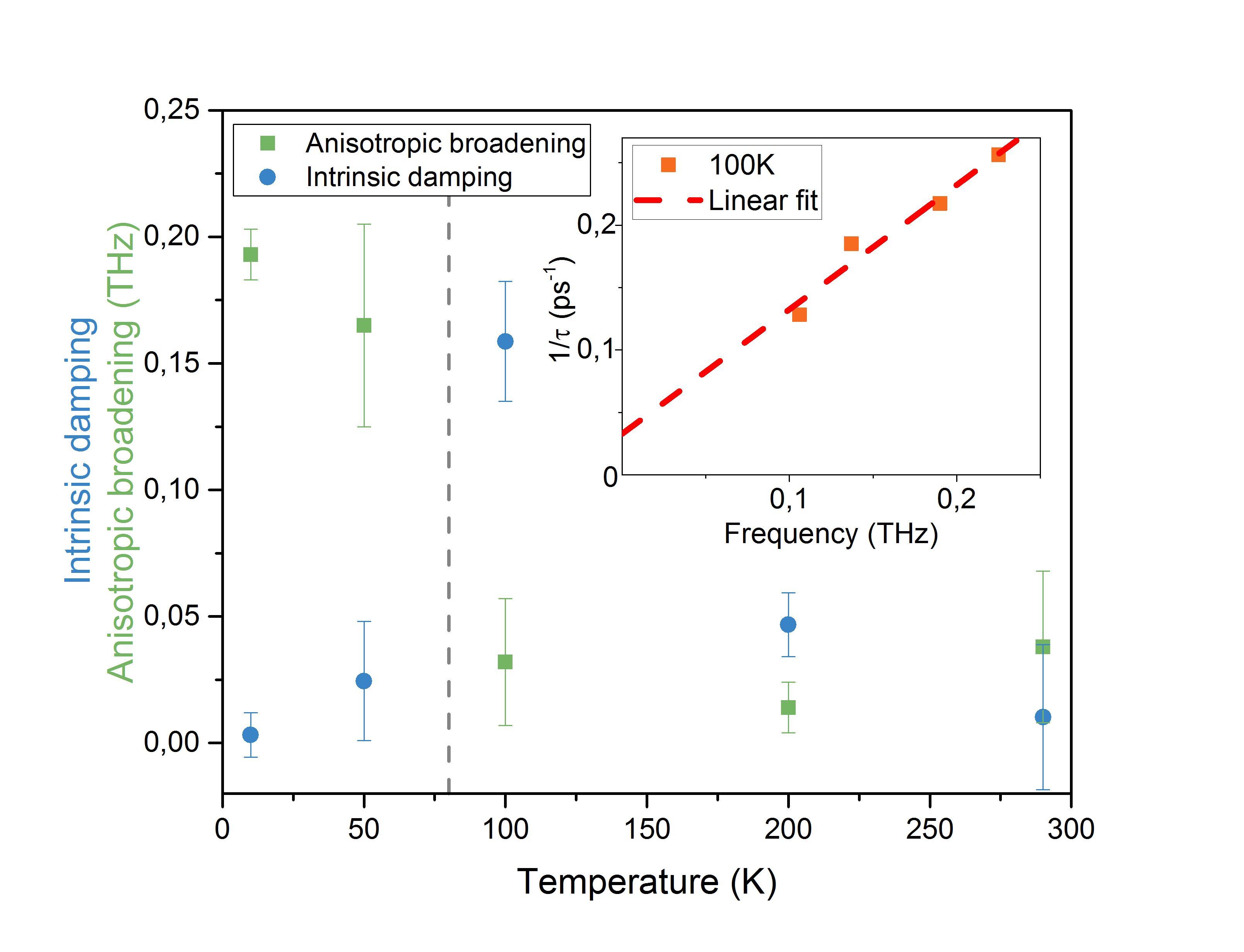}
  \caption{Intrinsic and anisotropic broadening in $\mrg$ across the $\tcmp$. The inset shows the evaluation process of the two damping parameters. A linear fit is used to evaluate intercept (anisotropic broadening) and slope (intrinsic damping) of the frequencies versus the inverse of the decay time. The data point are obtained from the fit of time-resolved Faraday effect measurements (an example is shown in Fig.\ref{fig:fiveTffit}).}
  \label{fig:alpha}
\end{figure}
\figurename~\ref{fig:alpha} shows the intrinsic damping $\alpha$ and the
anisotropic broadening as a function of temperature. Anisotropic broadening is
usually attributed to a variation of the anisotropy field in the region probed
by the probe pulse\cite{Walowski2008}. For \mrg\ this is due to slight lateral variations in the Ru content
$x$ in the thin film sample. Such a variation leads to a variation in
effective $\tcmp$ and $T_A$ and can therefore have a large influence on the
broadening as a function of temperature. Despite this, the anisotropic
broadening is reasonably low in the entire temperature range above $\tcmp$, and
a more likely explanation for its rapid increase below $\tcmp$ is that the
applied magnetic field is insufficient to completely remagnetize the sample
between two pump pulses. As observed in Fig.\ref{fig:geffandhk}, the anisotropy
field reaches almost \SI{4}{\tesla} at low temperature, comparable to our
maximum applied field of \SI{7}{\tesla}. The intrinsic damping $\alpha$ is
less than 0.02 far from $\tcmp$, but increases sharply at $T =
\SI{100}{\kelvin}$. We tentatively attribute this to an increasing portion of
the available power being transferred into the high-energy exchange mode, although we underline that we have not seen any direct evidence of such a mode in any of the experimental data.

\section{Conclusion}
\label{sec:conclusion}
We have shown that the time-resolved Faraday effect is a powerful tool to determine
the spin dynamic properties in compensated, metallic ferrimagnets. The high
spin polarisation of \mrg\ enables meaningful Faraday data to be recorded even near $\tcmp$ where the net magnetisation is vanishingly small, and the dependence
of the dynamics on the sublattice as opposed to the net magnetic properties
provides a more physical understanding of the material.
Furthermore, we find that the ferromagnetic-like mode of \mrg\ reaches resonance frequencies as high as $\SI{160}{\giga\hertz}$ in zero applied field, together with a small intrinsic damping. This value is remarkable if compared to well-known materials such as GdFeCo which, at zero field, resonates at tens of GHz\cite{Stanciu2006} or $[\text{Co/Pt}]_{n}$ multilayers at \SI{80}{\giga\hertz}\cite{Barman2007} but with higher damping. We should however stress that, in the presence of strong anisotropy fields, higher frequencies can be reached. Example of that can be found for ferromagnetic Fe/Pt with $\approx \SI{280}{\giga\hertz}$ ($H_{a}=10T$)\cite{Becker2014}, and for Heusler-like ferrimagnet (Mn$_{3}$Ge and Mn$_3$Ga) with $\approx\SI{500}{\giga\hertz}$ ($H_{a}=20T$)\cite{Mizukami2016,awari2016narrow}. Nevertheless, the examples cited above show a considerably higher intrinsic damping compared to \mrg. 
In addition, it was recently shown that \mrg\ exhibits unusually strong intrinsic spin-orbit torque\cite{lenne2019giant}. Thus, taking into account the material parameters we have determined here, it seems likely it will be possible to convert a DC driven current into a sustained ferromagnetic resonance at $f=\SI{160}{G\hertz}$, at least. These characteristics make \mrg, as well as any future compensated half-metallic ferrimagnet, particularly promising materials for both spintronics and all-optical switching.

\begin{acknowledgments}
  This project has received funding from the NWO programme Exciting Exchange, the \emph{European Union's} Horizon 2020 research and innovation programme under grant agreement No 737038 `TRANSPIRE', and from Science Foundation Ireland through contracts 12/RC/2278 AMBER and 16/IA/4534 ZEMS.
  
  The authors would like to thank D. Betto for help extracting $\langle L \rangle$ and $\langle S \rangle$.
\end{acknowledgments}

\bibliography{refarticle}
\end{document}